\documentclass[%
reprint,
superscriptaddress,
nofootinbib,
amsmath,
amssymb,
aps,
floatfix,
showkeys,
]{revtex4-2}

\usepackage[colorlinks,citecolor=blue,linkcolor=blue,anchorcolor=blue,filecolor=blue,
urlcolor=blue]{hyperref}

\usepackage{graphicx,color,xcolor}
\usepackage{amsfonts,amsmath}
\usepackage{amssymb,ulem}
\usepackage{dcolumn}
\usepackage{bm,lipsum}
\usepackage[utf8]{inputenc}
\usepackage{subeqnarray}
\usepackage{array}
\usepackage{epsfig,hyperref}
\usepackage{morefloats}
\usepackage{relsize}
\usepackage{url}
\usepackage{orcidlink}

\usepackage{comment}

\begin{document}

\preprint{ }

\title{Correlation between the symmetry energy slope and the deconfinement phase transition}

\author{Luiz L. Lopes$^1$, Debora P. Menezes$^2$, 
Mateus R. Pelicer$^3$}

\affiliation{$1$ Centro Federal de Educac\~{a}o Tecnol\'{o}gica de Minas Gerais Campus VIII; CEP 37.022-560, Varginha - MG - Brasil\\
$2$ Universidade Federal de Santa Catarina;  C.P. 476, CEP 88.040-900, Florianópolis, SC, Brasil \\
$3$ Department of Physics, Kent State University, Kent, OH 44243 USA}
% \mbox{$^2$Depto de F\'{\i}sica - CFM - Universidade Federal de Santa Catarina  Florian\'opolis - SC - CP. 476 - CEP 88.040 - 900 - Brazil }}

\date{\today}

\begin{abstract}
 We study how the nuclear symmetry energy slope ($L$) can affect the hadron-quark phase transition and neutron star properties. We show that the main physical quantities as the critical chemical potential and pressure are strongly influenced by the symmetry energy slope. In extreme cases, the total amount of deconfined quarks can reach up to 99$\%$ of the hybrid star mass.
\end{abstract}

\maketitle

\section{Introduction}

From terrestrial nuclear physics to neutron star observations, our knowledge about the behavior of strongly interacting matter has taken a great leap forward in the last 30 years. However, 
%up to today, there are still some open questions to be answered.
many open questions remain.
 
One of %the open questions
them is related to the neutron stars composition, for example, %interior. 
is there exotic, non-nucleonic matter in the neutron star core?
What kind? Some studies indicate that hyperons are inevitable~\cite{Dapo2010,lopesnpa,Tolos2017,lopes2023ptep}.
Other possibilities are the presence of $\Delta$ resonances~\cite{Tolos2019,MOTTA2020,lopesPRD} and kaon condensate~\cite{kaonp}.
An even more exotic scenario happens when the neutron star's inner core experiences a deconfinement phase transition and becomes composed of quarks, while the outer layers remain made up of hadronic matter. A star with a quark core surrounded by hadronic matter is called a hybrid star~\cite{Lopes_2022,hyb1,hyb2}. The study of hybrid stars' properties is crucial, as a recent study indicates that hybrid stars are indeed the most probable scenario for massive neutron stars~\cite{nature_2020}.

Another open question is the value of the symmetry energy slope. In the earlier 2010s, most studies pointed to a relatively low value for $L$. For instance, in refs.~\cite{Paar2014,Steiner2014,Lattimer2013}  upper limits of 54.6, 61.9, and 66 MeV respectively were suggested. However, in the last couple of years, the situation has changed and new experiments have pointed to a significantly higher upper limit.
For instance, in a study about the spectra of pions in intermediate energy collisions, an upper limit of 117 MeV was obtained~\cite{pions}, while in one of the PREX II analyses~\cite {PREX2} an upper limit of 143 MeV was suggested.
All these conflicting results have been well summarized in a recent paper (ref.~\cite{Tagami2022}): the CREX group points to a slope in the range 0 $<~L~<$ 51 MeV, while PREX II results point to 76 MeV $<~L~<$ 165 MeV. Although, a revised analysis of PREX II data, including also ground state properties of nuclei,  astrophysical observations, and heavy-ion collision indicates a slope in the range 59 $<~L~<$ 107 MeV~\cite{Yue2022} the verdict does not change: there is no overlap between CREX and PREX II analyses. This is a big 
%problem to be solved. 
caveat.

In this work, we assume that every supermassive neutron star must have a quark core, as suggested in ref.~\cite{nature_2020}, and study how different values of the symmetry energy slope $L$ affect their properties. We begin with microscopic properties. For different values of $L$, we construct the equation of state (EoS), and then pay special attention to the physical quantities related to the phase transition, such as the point at which the Gibbs free energy per baryon $G/n_B$  of both phases intersect (commonly called critical chemical potential $\mu_0$~\cite{Chamel2013}), the critical pressure ($p_0$), the energy density gap between the two phases, ($\Delta\epsilon$), the latent energy ($\Delta Q$)~\cite{latent,mateus}, and the hadronic number density $n$.
%I changed latent heat to latent energy, because we are studying zero temperature, so in principle, there is no heat. Someone discussed this with me because I called it heat in the other paper.

We then study how $L$ affects the macroscopic properties of the neutron stars, as their maximum mass, the radius of the canonical 1.4 M$_\odot$ star, the minimum star mass that allows a quark core ($M_{min}$) and the total mass and size of the quark core inside the hybrid star. To accomplish these tasks we use an extended version of quantum hadrodynamics (QHD)~\cite{Serot_1992} for the hadronic phase and a thermodynamic consistent vector MIT model for the quark phase, both in the mean-field approximation~\cite{lopesps1}.

In the case of the QHD, besides the traditional $\sigma\omega\rho$ mesons, we use two extensions: to
reduce the slope, we add the non-linear $\omega \rho$ coupling as presented in the IUFSU model~\cite{Rafa2011,IUFSU,dex19jpg}; while to increase the slope, we add the scalar-isovector $\delta$ meson~\cite{KUBIS1997,Liu2002,Lopes2014BJP}.  The use of two different interactions helps us understand both the phenomenology of changes in the slope $L$ and the influence of different fields and coupling constants in the field theory.
In the case of the vector MIT bag model, we use symmetry group arguments to fix the coupling constant of the vector field with different quark flavors~\cite{lopesps1}.

\section{Formalism}

The extended version of the QHD~\cite{Serot_1992}, which includes both the $\omega\rho$ non-linear coupling~\cite{IUFSU,Rafa2011,dex19jpg} and the scalar-isovector $\delta$ meson~\cite{KUBIS1997,Liu2002,Lopes2014BJP}, has the following Lagrangian density in natural units:
\begin{eqnarray}
\mathcal{L}_{QHD} =  \bar{\psi}_N[\gamma^\mu(\mbox{i}\partial_\mu  - g_{\omega}\omega_\mu   - g_{\rho} \frac{1}{2}\vec{\tau} \cdot \vec{\rho}_\mu)+ \nonumber \\
- (M_N - g_{\sigma}\sigma - g_{\delta}\vec{\tau} \cdot \vec{\delta})]\psi_N  -U(\sigma) +   \nonumber   \\
  + \frac{1}{2}(\partial_\mu \sigma \partial^\mu \sigma - m_s^2\sigma^2)  + \frac{1}{2}(\partial_\mu \vec{\delta} \cdot \partial^\mu \vec{\delta} - m_\delta^2\delta^2) + \nonumber \\ - \frac{1}{4}\Omega^{\mu \nu}\Omega_{\mu \nu} + \frac{1}{2} m_v^2 \omega_\mu \omega^\mu+ \Lambda_{\omega\rho}(g_{\rho}^2 \vec{\rho^\mu} \cdot \vec{\rho_\mu}) (g_{\omega}^2 \omega^\mu \omega_\mu) + \nonumber \\
 + \frac{1}{2} m_\rho^2 \vec{\rho}_\mu \cdot \vec{\rho}^{ \; \mu} - \frac{1}{4}\bf{P}^{\mu \nu} \cdot \bf{P}_{\mu \nu}  , \label{s1} 
\end{eqnarray}
% in natural units. 
 The $\psi_N$  is the  Dirac field of the nucleons.  The $\sigma$, $\omega_\mu$, $\vec{\delta}$ and $\vec{\rho}_\mu$ are the mesonic fields.
 The $g's$ are the Yukawa coupling constants that simulate the strong interaction, $M_N$ is the nucleon mass and  $m_s$, $m_v$, $m_\delta$ and $m_\rho$ are
 the masses of the $\sigma$, $\omega$, $\delta$ and $\rho$ mesons respectively.
The $U(\sigma)$ is the self-interaction term introduced in ref.~\cite{Boguta} to fix the compressibility:

\begin{equation}
U(\sigma) =  \frac{\kappa M_N(g_{\sigma} \sigma)^3}{3} + \frac{\lambda(g_{\sigma}\sigma)^4}{4} \label{sbog} .
\end{equation} 

Furthermore, leptons are added as free fermions to account for the chemical stability. The EoS is then obtained in mean field approximation (MFA) by calculating the components of the energy-momentum tensor~\cite{Serot_1992}. The total energy density and the number density are given by~\cite{Miyatsu2013,Lopes2014BJP}:

\begin{flalign}
 \epsilon = \sum_N &\frac{1}{\pi^2}\int_0^{k_{Nf}} dk k^2 \sqrt{k^2 + M_N^{*2}} +U(\sigma_0)  \nonumber \\
 &+\frac{1}{2}m_\sigma^2\sigma_0^2 + \frac{1}{2}m_\omega^2\omega_0^2 + 
 \frac{1}{2}m_\delta^2\delta_0^2 
 + \frac{1}{2}m_\rho^2\rho_0^2  \nonumber \\
  &+3 \Lambda_v\omega_0^2\rho_0^2 
  \label{EL7}  + \sum_l \frac{1}{\pi^2}\int_0^{k_{lf}} dk k^2 \sqrt{k^2 + m_l^{2}} , \\
  n = \sum_B &\frac{k_{f_N}^3}{3\pi^2}
\end{flalign}
where $\Lambda_v = \Lambda_{\omega\rho}g_\omega^2g_\rho^2$; $N$ indicates the nucleon
and $l$ indicates leptons. The pressure is easily obtained by thermodynamic relations: $p = \sum_f \mu_f n_f - \epsilon$, where the sum runs over all the fermions and $\mu_f$ is the corresponding chemical potential.

The parameters used in this work are based on the  L3$\omega\rho$ parametrization introduced in ref.~\cite{Lopes2022CTP} and are presented in Tab.~\ref{TL1}. The nuclear constraints at the saturation density are also in Tab.~\ref{TL1} and are taken from two extensive review articles, ref.~\cite{Dutra2014,Micaela2017}. The parameters $(g_\rho/m_\rho)^2$, $(g_\delta/m_\delta)^2$  and $\Lambda_{\omega\rho}$ are chosen in order to fix the symmetry energy at the saturation point $S_0$ = 31.7 MeV, while varying the slope $L$. Their values are presented in Tab.~\ref{T2}.  It is worth pointing out that fixing the symmetry energy at the saturation density is not the only possibility. In ref.~\cite{Horo2001} the authors fix $S$ at $k_f$ = 1.15 fm ($n ~\approx~0.1$ fm$^{-3}$) instead.

\begin{table}[tb]
\begin{center}
\begin{tabular}{cc|cccc}
\toprule
  & Parameters & &  Constraints  & Our model  \\
\toprule
 $\left(g_{\sigma}/{m_s}\right)^2$ & $12.108 \, \mathrm{fm}^2$ &$n_0 (\mathrm{fm}^{-3})$ & 0.148 - 0.170 & 0.156 \\
% \hline
$\left(g_{\omega}/{m_v}\right)^2$  & $7.132 \, \mathrm{fm}^2$ & $M^{*}/M$ & 0.6 - 0.8 & 0.69  \\
%  \hline
  $\kappa$ & 0.004138 & $K \mathrm{(MeV)}$ & 220 - 260             &  256  \\
% \hline
$\lambda$ &  -0.00390  & $S_0 \mathrm{(MeV)}$  & 28.6 - 34.4 &  31.7  \\
%\hline
- &  - & $B/A \mathrm{(MeV)}$  & 15.8 - 16.5  & 16.2 \\
\toprule
\end{tabular}
\caption{Model parameters used in this study and their predictions for symmetric nuclear matter at saturation density. The phenomenological constraints were taken from Refs.~\cite{Dutra2014, Micaela2017}.}
\label{TL1}
\end{center}
\end{table}

%------------------------------------------------------
\begin{center}
\begin{table}[bt]
\begin{center}
\begin{tabular}{ccccc}
\toprule
  $L$ (MeV) & $(g_\rho/m_\rho)^2$ ($\mathrm{fm}^2$) & $(g_\delta/m_\delta)^2$ ($\mathrm{fm}^2$)  & $\Lambda_{\omega\rho}$  \\
\toprule
 44 & 8.40 & 0  & 0.0515 \\
 %\hline
 60 & 6.16  & 0  & 0.0344 \\
%  \hline
 76& 4.90 & 0 & 0.0171        \\
% \hline
92 &  4.06  & 0 & 0  \\
%\hline
100 &  7.23 & 0.92 & 0   \\
%\hline 
108 &  10.41 & 1.85 & 0   \\
%\hline 
116 &  13.48 & 2.76 & 0   \\
\toprule
\end{tabular}
\caption{Model parameters selected to set the symmetry energy at $S_0$ = 31.7 MeV.} 
\label{T2}
\end{center}
\end{table}
\end{center}

The parameters used in this work are the same as the ones presented in ref.~\cite{lopescesar}, where the authors vary the symmetry energy slope to study pure hadronic neutron stars and gravitational wave signals. Additional discussions about the  QHD formalism, calculations, behavior of the symmetry energy and its slope,  as well as the role played by the $\rho$, $\delta$ mesons, and the non-linear couplings can be found in ref.~\cite{lopescesar} and references therein.

Now, the thermodynamic consistent vector MIT bag model has the following Lagrangian density~\cite{lopesps1,Lopes_2022,Lopes2022ApJ}:
\begin{eqnarray}
\mathcal{L}_{vMIT} = \{ \bar{\psi}_q[\gamma^\mu(i\partial_\mu  - g_{qV}V_\mu)  - m_q ]\psi_q \nonumber \\
+ \frac{1}{2}m_V^2V_\mu V^\mu - B \}\Theta(\bar{\psi}_q\psi_q) . \label{s2}
\end{eqnarray}
where $m_q$ is the mass of the quark $q$ of flavor $u$, $d$ or $s$.
Here, we follow ~\cite{lopesps1} and use ($m_u=m_d=4$ MeV,
$m_s=95$ MeV); $\psi_q$ is the Dirac quark 
field, $B$ is the constant vacuum pressure, and $\Theta(\bar{\psi}_q\psi_q)$ is the Heaviside step function to assure that the quarks exist only confined to the bag.
The quark interaction is mediated by the massive vector channel
$V_\mu$ analogous to the $\omega$ meson in QHD~\citep{Serot_1992}.  Again,
leptons are added to account for $\beta$ stable matter.  The construction of the EoS and the computation of the related physical quantities are analogous to the QHD calculations, i.e., by applying the Euler-Langrage equations in eq.~\ref{s2}, and relying on the MFA we obtain the energy eigenvalue~\cite{lopesps1,Lopes_2022}:

\begin{equation}
  E_q = \sqrt{k^2 + m_q^2} + g_{qV}V_0 . \label{qee}
\end{equation}

 The total energy density and the number density for the quarks are given by:
\begin{flalign}
 \epsilon = \sum_q &\frac{N_c}{\pi^2}\int_0^{k_{fq}} dk k^2 \sqrt{k^2 + m_q^{2}} + \frac{1}{2}m_V^2V_0^2 
\nonumber \\    &+ B  + \sum_l \frac{1}{\pi^2}\int_0^{k_{lf}} dk k^2 \sqrt{k^2 + m_l^{2}}\\
  n = \sum_q & N_c \times \frac{k_{fq}^3}{3\pi^2} , 
\end{flalign}
 where $N_c = 3$ is the number of colors. The pressure is again obtained by thermodynamic relations: $p = \sum_f \mu_f n_f - \epsilon$. the bag $B$ is taken as $B^{1/4}$ = 158 MeV, as discussed in ref.~\cite{Lopes_2022}. Two other important quantities are $X_V$ and $G_V$ defined as~\cite{lopesps1,Lopes_2022}
\begin{equation}
X_V = \frac{g_{sV}}{g_{uV}}, \quad G_V = \bigg ( \frac{g_{uV}}{{m_V}} \bigg )^2. \label{s3}
\end{equation}
$X_V$ is related to the relative strength of the vector field with the $s$ quark in relation to the $u$ and $d$ quarks. We assume $X_V$ = 0.4, since this is the value predicted by the symmetry group~\cite{lopesps1}. $G_V$ is related to the absolute strength of the vector field, and we use here three different values, 0.38, 0.40 and 0.42 fm$^2$ to guarantee that the quark hadron phase transition happens for all values of $L$.   It is important to notice that varying the bag constant ($B$) causes a global effect on the quark EoS (the entire EoS is shifted). Increasing the bag causes a softening in the EoS, reducing the critical chemical potential and the maximum mass of the related compact object. On the other hand, reducing the bag causes a stiffening of the EoS. However, a larger reduction in the bag can make the quark matter energetically unfavorable and prevent the formation of the hybrid star. Now, if we change $X_V$ to 1.0 (a universal coupling) it stiffens the EoS, but in general, it will require further modifications on the values of $G_V$, otherwise, the quark EoS will be energetically unfavorable.  Additional discussions on the role of $B$, $X_V$, and $G_V$ can be seen in ref.~\cite{Lopes_2022,Lopes2022ApJ}. Moreover, with the values of $B$, $X_V$, and $G_V$ used in this work, we guarantee that the quark matter is outside the so-called stability window~\cite{Bodmer,Witten}. 
It was shown in ref.~\cite{Olinto} that if the quark matter in the core of hybrid stars were obtained with parameters that lie inside the stability window (absolutely stable strange matter), the entire star would be converted into a strange star in a finite amount of time.

\section{Results and discussion}

\subsection{Microscopic results}

We use the so-called Maxwell construction for the deconfinement transition. In this approach,  the transition pressure is the one
where the Gibbs free energy per baryon $G/n_B$ of both phases intersect, being the energetically preferred phase the one with lower $G/n_B$~\cite{Chamel2013}. The Gibbs free energy per baryon coincides with the baryon chemical potential. We call the intersection points critical pressure ($p_0$) and critical chemical potential ($\mu_0$). The chemical potential at T = 0 for hadron and quark phases can be calculated as~\cite{mateus,Bombaci}:
\begin{figure}
  \begin{centering}
\begin{tabular}{c}
\includegraphics[width=0.333\textwidth,angle=270]{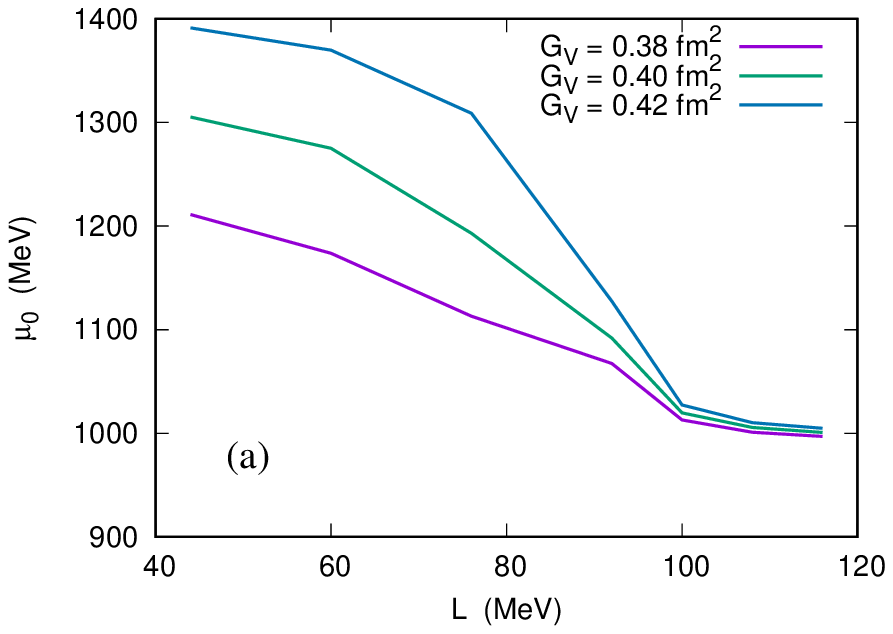} \\
\includegraphics[width=0.333\textwidth,,angle=270]{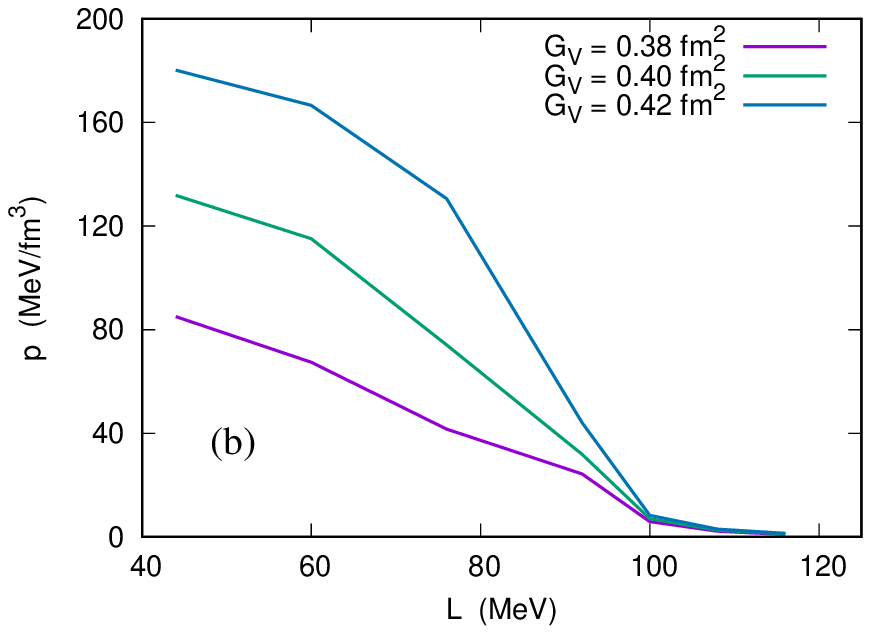} \\
\end{tabular}
\caption{Critical chemical potential (a) and pressure (b) as a function of the symmetry energy slope $L$, for different quark-meson couplings: $G_V=$ 0.38, 0.40, and 0.42 fm$^2$. (Color online).} \label{Ftr}
\end{centering}
\end{figure}
\begin{equation}
 \mu_H =  \frac{\epsilon_H + p_H}{n_H}, \quad \mu_Q = \frac{\epsilon_Q + p_Q}{n_Q}   ,
\end{equation}
and the criteria for the Maxwell construction are:
\begin{equation}
\mu_0 = \mu_H = \mu_Q, \quad \mbox{and} \quad p_0 =  p_H =  p_Q.
\end{equation}

So far, there are no experimental results that could indicate 
the critical chemical potential  at T = 0, and our current knowledge relies on effective models.   Here, we follow ref.~\cite{lopesps2} that suggests that the critical chemical potential is in the range 1050 MeV $<~\mu_0~<$ 1400 MeV.

\begin{figure}
  \begin{centering}
\begin{tabular}{c}
\includegraphics[width=0.333\textwidth,angle=270]{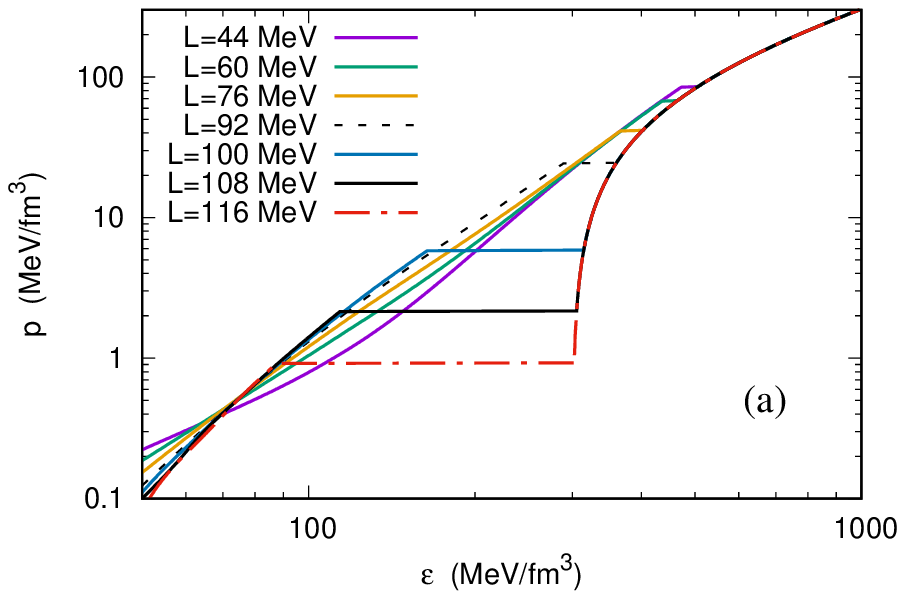} \\
\includegraphics[width=0.333\textwidth,angle=270]{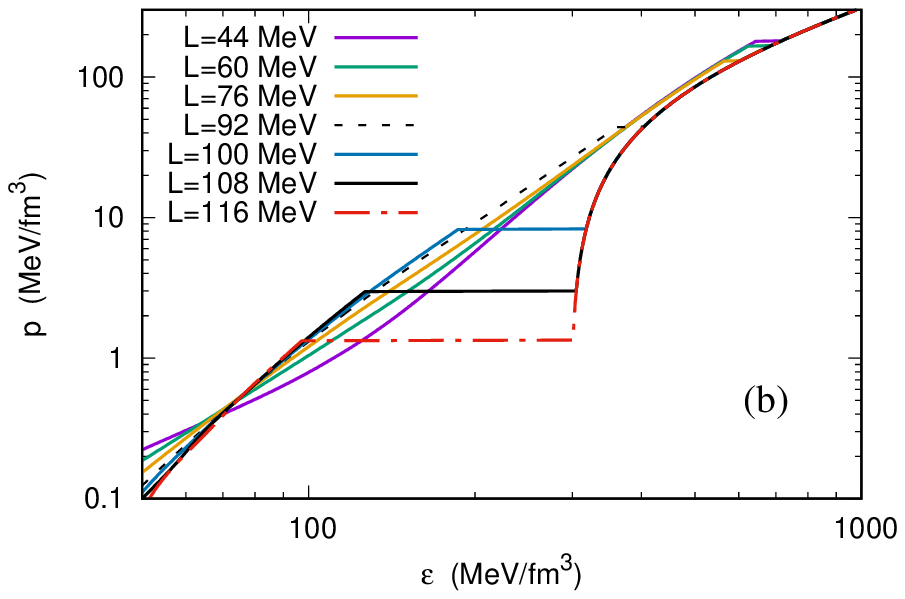} \\
\end{tabular}
\caption{Hybrid EoS for different slope values of the hadron phase and for quark vector couplings $G_V=$0.38 (a) and 0.42 fm$^2$ (b) in the top and bottom figures, respectively. (Color online)} \label{Feos}
\end{centering}
\end{figure}

In Fig.~\ref{Ftr} we shown the critical chemical potential and the critical pressure in function of the slope $L$ for three values of $G_V$. We can see that for all values of $G_V$, both, $\mu_0$ and $p_0$ decrease with increasing slope. Thus, for all values of $G_V$ used, the higher critical values are obtained for the smaller slopes. We also notice that when we assume $L~\geq$ 100 MeV, both critical chemical potential and pressure become almost independent of the value of $G_V$.  Moreover, the critical chemical potential drops below 1050 MeV, which strongly disfavors very high values of $L$ as the constraint from  PREX II results~\cite{PREX2}. This can be better understood from the hybrid EoS curves in  Fig.~\ref{Feos}: a larger slope implies a stiffer hadronic EoS in the deconfinement region, so the critical pressure is reached earlier. At lower pressures ($p\leq 0.4~$MeV~fm$^{-3}$), the behavior of the EoS with the slope is reversed, and the larger slope corresponds to a softer EoS. { This occurs because of the strength and nature of the coupling constant, as discussed in ref.~\cite{lopescesar}. Nevertheless, it does not affect our results, since there is no deconfinement in this region.

\begin{figure}
  \begin{centering}
\begin{tabular}{c}
\includegraphics[width=0.333\textwidth,angle=270]{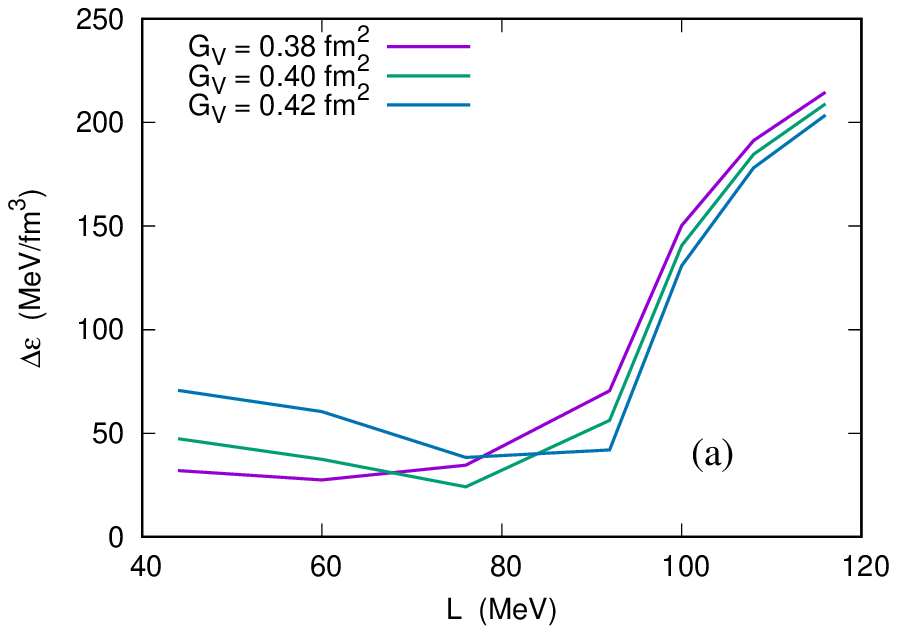} \\
\includegraphics[width=0.333\textwidth,,angle=270]{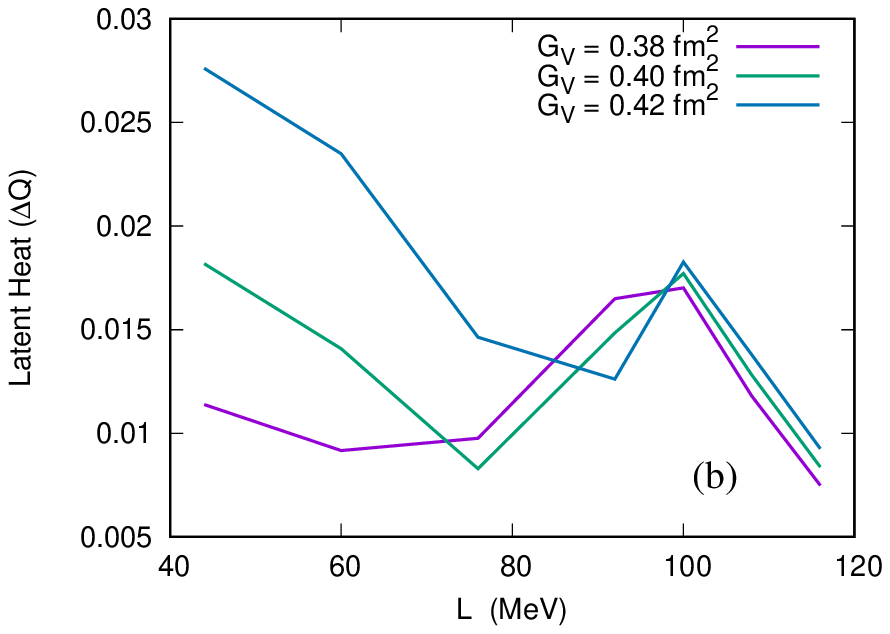} \\
\end{tabular}
\caption{ Energy difference (a) and latent energy (b) of the deconfinement transition as a function of the slope. Each curve corresponds to a different quark meson coupling, $G_V=$0.38, 0.40 and 0.42 fm$^2$. (Color online). } \label{Fq}
\end{centering}
\end{figure}
An interesting physical quantity is the discontinuity of the energy density, or energy density gap, { a feature that appears due to the Maxwell construction choice}: $\Delta \epsilon =  \epsilon_Q - \epsilon_H$.  For low values of $L$, the values of $\Delta \epsilon $ present little change, and are always in the range  25 MeV/fm$^3$ $<~\Delta \epsilon ~<$ 70  MeV/fm$^3$. However, as the slope increases, so does the discontinuity of the energy density. And here again, the results become almost independent of $G_V$. For extreme cases ($L = 116$ MeV), the value of $\Delta \epsilon $ always surpasses 200 MeV/fm$^3$.  This is illustrated in the top plot of fig.~\ref{Fq}, which shows the energy difference between quark and hadron phases as a function of the slope.

At the bottom of Fig.~\ref{Fq} we show the relativistic (adimensional) latent energy 
\begin{equation}
    \Delta Q = p_0 \frac{\varepsilon_Q- \varepsilon_H}{ \varepsilon_H \varepsilon_Q},
\end{equation}
which was proposed in ref.~\cite{latent} as an intuitive generalization of the non-relativistic latent heat (see eqs~(2) and~(4) of~\cite{latent}). It should be mentioned that this quantity is not unique in determining the phase transition discontinuity,  and different quantities have been defined for this purpose in the literature~\cite{seidov,PhysRevC.83.024308,PhysRevD.58.024008}.  Still, this is an interesting quantity, as we comment next. 
%as upper bounds for it were established in ref.~\cite{latent} using pQCD.  
In our calculation, the larger values of the latent energy are obtained for the smallest slope considered, of 44~MeV. Furthermore, in general, the higher the value of $G_V$, the higher is the latent heat. This is due to the large critical pressure and energy densities where deconfinement occurs. Afterwards, $\Delta Q$ decreases until around $80 \lesssim L\lesssim 90~$MeV, and a local maximum occurs around $L\approx 100$~MeV, which is the point where there is a steep decrease in the critical pressure and energy densities of both phases, but an increase in the energy density discontinuity. We should expect $\Delta Q\rightarrow 0$ if we keep increasing the slope, as the critical pressure drops quickly as already shown in Fig.~\ref{Ftr}.

\begin{center}
\begin{table}%[ht]
\begin{center}
\scalebox{0.92}{
\begin{tabular}{cc|cccccc}
\toprule
$G_V$ (fm$^2$) &  $L$ (MeV) & $\mu_0$ (MeV) & $p_0$   & $\Delta \epsilon$  & $\Delta$ Q & $n_c/n_0$\\
\toprule
0.38& 44 & 1211  & 85.0  & 32.0 & 0.0113 & 2.95 \\
 %\hline
0.38& 60 &  1174  & 67.4  & 27.5 & 0.0091 & 2.73  \\
  %\hline
0.38& 76& 1112  & 41.5 & 34.6   & 0.0097  & 2.36   \\
 %\hline
0.38& 92 & 1067   & 24.2 & 70.5 & 0.0164 & 1.88 \\
%\hline
0.38& 100 & 1012  & 5.8& 150.2  & 0.0170 & 1.09  \\
%\hline 
0.38& 108 & 1000   & 2.1 & 191.3 & 0.0118 & 0.77 \\
%\hline 
0.38& 116 & 996  & 0.92 & 214.6  & 0.0074 & 0.60 \\
%\hline 
 \hline
 0.40& 44 & 1305 & 131.8  & 47.4 & 0.0181  & 3.41\\
 %\hline
 0.40&60 & 1275  & 115.1  & 37.4 & 0.0140 & 3.26\\
 % \hline
 0.40&76& 1193 & 74.1 & 24.1   & 0.0082 & 2.83   \\
 %\hline
0.40&92 & 1091   & 31.8 & 56.2 & 0.0148  & 2.06 \\
%\hline
0.40&100 & 1019  & 6.9 & 140.6 & 0.0177 & 1.16  \\
%\hline 
0.40&108 &  1005 & 2.5 & 184.6  & 0.0128 & 0.81  \\
%\hline 
0.40&116 & 1000  & 1.1 &  209.0 & 0.0083 & 0.63 \\
%\hline
 \hline
 0.42&44 & 1391  & 180.2  & 70.7 &  0.0276 & 3.79\\
% \hline
 0.42&60 & 1369  & 166.6  & 60.4 & 0.0234 & 3.70 \\
 % \hline
 0.42&76& 1308  & 130.5 &  38.3  & 0.0143 & 3.40  \\
 %\hline
0.42&92 &  1127  & 44.0 & 41.9 & 0.0126 & 2.30 \\
%\hline
0.42&100 & 1027  & 8.3 & 130.8 & 0.0182 & 1.25 \\
%\hline 
0.42&108 & 1010  & 2.9 & 178.1 & 0.0137 & 0.85  \\
%\hline 
0.42&116 & 1004  & 1.3 & 203.5  & 0.0092 & 0.66 \\
\toprule
\end{tabular}}
 
\caption{ Some physical quantities at the quark-hadron phase transition for different values of $L$ and $G_V$.} 
\label{T3}
\end{center}
\end{table}
\end{center}

Another interesting quantity is the hadronic number density at the phase transition, $n_c$. Although there is no formal bond on this parameter, there are some reasonable statements that we can make. First, the quark-hadron phase transition is very unlikely to happen below 1.3 times the nuclear saturation density, $n_0$,  because, as pointed in ref.~\cite{latent}, the EoS up this value can be modeled by next-to next-to next-to-leading order from chiral perturbation theory (ChPT). Indeed, 1.3 $n_0$ was also used as a lower limit in the study of latent heat presented in ref.~\cite{latent}.
Ultimately, we can consider unrealistic a model that predicts the quark-hadron phase transition at densities lower than the nuclear saturation density itself.
In tab.~\ref{T3} we summarize the critical pressure and chemical potential, the energy density discontinuity,  the latent energy, and the hadronic number density for different slopes and quark-meson couplings. As can be seen, the constraint $\mu_0~>$ 1050 MeV suggested in ref.~\cite{lopesps2} and the bound $n_c~>1.3~n_0$ are in accordance.
Moreover, for $L~\geq$ 108 MeV, the results become even more extreme. We have $\mu_0$ $\approx$ 1000 MeV and the number density of the phase transition drops below the nuclear saturation density. These results strongly disfavor larger values of the slope. From the field theory point of view, our results also strongly advocate against the inclusion of the scalar-isovector $\delta$ meson in the standard $\sigma\omega\rho\delta$ model of the QHD if there is no $\omega\rho$-interaction~\cite{KUBIS1997,Liu2002}.

\subsection{Macroscopic Results}

\begin{figure}[b]
  \begin{centering}
\begin{tabular}{c}
\includegraphics[width=0.333\textwidth,angle=270]{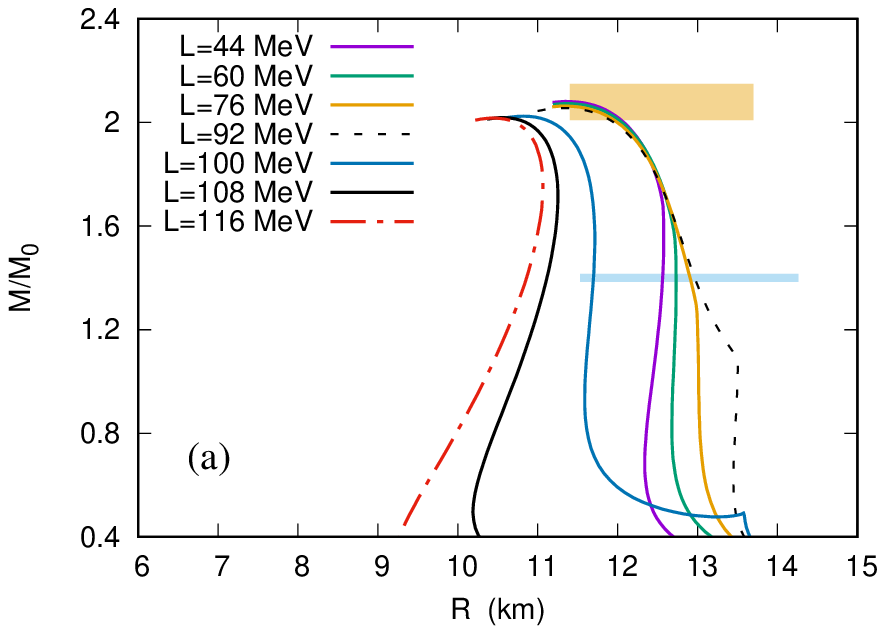} \\
\includegraphics[width=0.333\textwidth,,angle=270]{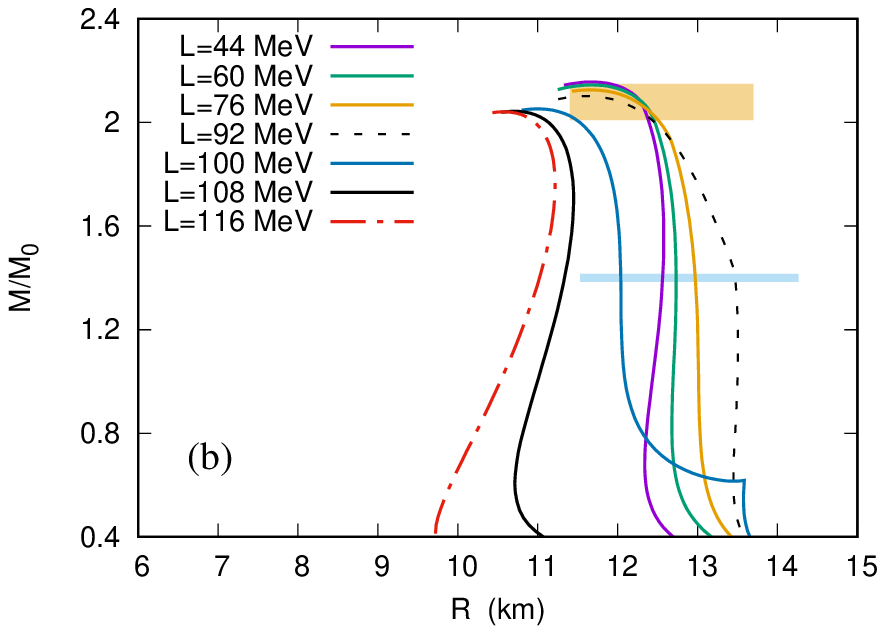} \\
\end{tabular}
\caption{Mass radius diagram for hybrid stars within $G_V=0.38$ (a) and $G_V=0.42~$fm$^2$ (b). Different curves represent different slope values for hadron matter.  The hatched areas are the constraints coming from the  PSR J0740+6620~\cite{Riley2021} and the radius of the canonical star as discussed in the text. (Color online)}  \label{Ftov}
\end{centering}
\end{figure}

Now we turn to analyze how the change in slope modifies the mass-radius diagram for hybrid stars. For each of the discussed EoS we solve the TOV equations~\cite{TOV} and obtain the mass-radius diagram. Moreover, we use the BPS EoS~\cite{BPS} for the neutron star's outer crust and the BBP EoS~\cite{BBP} for the inner crust.
In fig.~\ref{Ftov} we show the influence of the slope  for $G_V=~$0.38 and 0.42~fm$^2$ in the top and bottom panels, respectively. We also discuss a couple of observational constraints coming from the NICER X-ray telescope. The first, and maybe the most important one, is the PSR J0740+6620, whose mass and radius lie in the range of $M = 2.08\pm 0.07~M_\odot$, and 11.41 km $<~R~<$ 13.70 km, respectively~\cite{Riley2021}.
The other constraint is related to the radius of the canonical $M = 1.4~M_\odot$ star.
Two NICER results constrain the radius of the canonical star between  11.52 km $<~R_{1.4}~<$ 13.85 km~\cite{Riley:2019yda} and between 11.96 and 14.26 km~\cite{Miller:2019cac}. Here we use the union set of both constraints as a more conservative approach. Explicitly, we use 11.52 km $<~R_{1.4}~<$ 14.26 km as a constraint. The results and constraints are presented in Fig.~\ref{Ftov}.

Let's begin by analyzing the effect of the slope $L$ for a fixed $G_V$. We notice that the increase of $L$ causes a decrease in the maximum mass that can reach around  0.1$M_\odot$. Moreover, as shown in Tab.~\ref{T3}, for higher values of $L$, the critical chemical potential is very low. This results in the minimum mass that presents a quark core ($M_{min})$ being as low as 0.1 $M_\odot$. As a consequence, for higher values of $L$, the  hybrid stars with masses around and above the canonical 1.4 $M_\odot$ present very low values of radii.
Indeed, we can see that for $L~\geq$ 108 MeV, all the radii for the canonical star fall below the lower limit of 11.52 km discussed above. The same is true for the radius constraint of the  PSR J0740+6620~\cite{Riley2021}. Although we have a mass in the $M = 2.08\pm 0.07~M_\odot$ range, the radii for $L~>100$ MeV are below the lower limit of 11.41 km.

In relation to $G_V$, we can see that increasing $G_V$ increases the maximum mass of the hybrid star. Also, since the critical chemical potential increases with a larger $G_V$,  the minimum mass that allows a quark core ($M_{min})$ also increases. In relation to the constraints, we see that we have a congruence in the bounds. The same results that predict a phase transition below the saturation density are the ones outside of the constraints of both the radius of the canonical star and the radius range of the PSR J0740+6620~\cite{Riley2021}. They also predict a critical chemical potential within $\mu_0~<$ 1050 MeV.
From the field theory point of view, all the results that fail to fulfill the supra-cited bonds, are those that employ the scalar-isovector $\delta$ meson. Indeed, the only results that use the $\delta$ meson and are still able to produce 11.52 km $<~R_{1.4}~<$ 14.26 km are those with $L = 100$ MeV. However, it still fails to reproduce the radii of the PSR J0740+6620. Therefore, the neutron star constraints also advocate against the scalar-isovector $\delta$ meson.

 In a recent work (ref.~\cite{Sun2023}), the authors suggested that the value of the energy gap, $\Delta\epsilon$ is related to the maximum mass of a hybrid star, such that the higher the gap, the lower the maximum mass. Indeed, they pointed out that if  $\Delta\epsilon~>$  180 MeV/fm$^3$, the maximum mass of the corresponding hybrid stars do not meet the lower mass limit of the second object in the GW190814 event. Our results support only partially this discussion. Although it is true that the higher value of $\Delta\epsilon$ produces the lower maximum mass, the opposite is not true, i.e., the lowest value of $\Delta\epsilon$ does not produce the most massive star. Moreover, even by increasing  $\Delta\epsilon$ by six or seven times, the reduction in the maximum mass is only around 3-5$\%$. A more natural correlation seems to be between the maximum mass and the critical chemical potential. The higher the value of the critical chemical potential, the higher the maximum mass.
 A similar result was already found in ref.~\cite{Lopes2022ApJ} with different models and parametrizations. The role of the energy gap seems to be therefore only secondary.

\begin{figure}[h]
  \begin{centering}
\includegraphics[width=0.333\textwidth,angle=270]{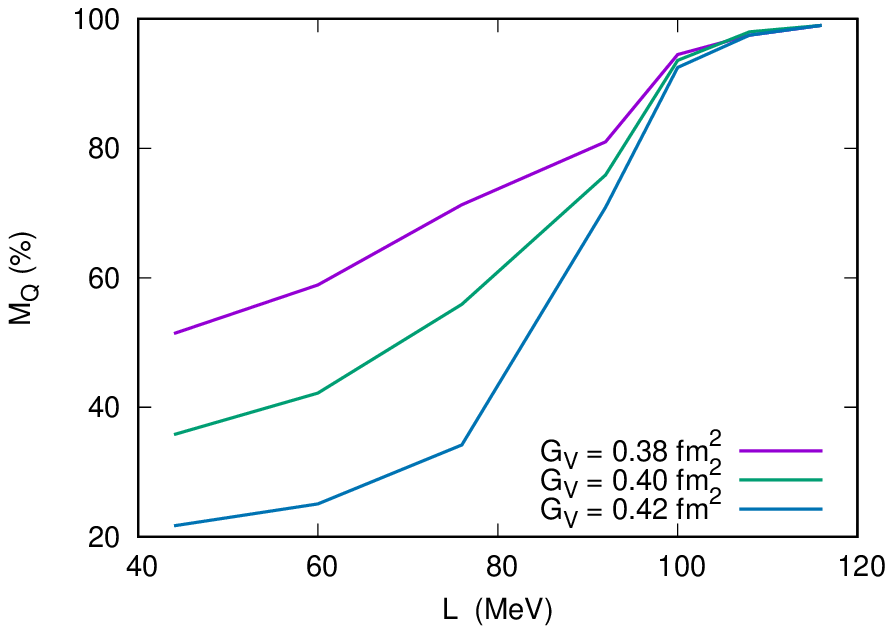} \\
\caption{ Percentual amount of quarks in the core of the most massive hybrid stars. The amount of quarks can be as high as $99\%$ of the total mass of the star} \label{Fmq}
\end{centering}
\end{figure}

\begin{widetext}
\begin{center}
\begin{table}[ht]
\begin{center}
\scriptsize
\begin{tabular}{cc|ccccccccccc}
\toprule
$G_V$ (fm$^2$) & $L$ (MeV) & $M_{m} (M_\odot)$ & R (km) &$n_{m}/n_0$   & $M_{min}$ ($M_\odot$) &$R_{1.4}$  (km) &  $M_{Q}$ ($M_\odot$) & $M_{Q}$ ($\%$) & $R_{Q}$  (km) &$R_{Q}$  ($\%$) &J0740+6620\\
 %\hline
0.38 & 44 & 2.08 & 11.34  & 6.30 & 1.67 & 12.58  & 1.07 & 51.4 & 7.35 & 64.7 & Yes\\
 %\hline
 0.38& 60 & 2.07  & 11.35  & 6.33 & 1.52  & 12.74 & 1.22  & 58.9  & 7.83 & 68.9& Yes\\
  %\hline
 0.38& 76& 2.06 & 11.35 & 6.35 & 1.28  & 12.90 & 1.47 &  71.3 & 8.58 & 75.5 & Yes\\
 %\hline
0.38&92 & 2.06  & 11.34 & 6.36 & 1.07 & 12.96 & 1.67 & 81.0 & 9.15 & 80.6 & Yes\\
%\hline
0.38& 100 & 2.02  & 10.82  & 6.65 & 0.49 & 11.70 & 1.91 &  94.5 & 9.80  & 90.5 &No\\
%\hline 
0.38& 108 & 2.02  & 10.51 & 6.86 & 0.23 &11.11  & 1.97 & 97.5 & 9.92  & 94.3 & No\\
%\hline 
0.38& 116 & 2.01  & 10.35 & 7.11 & 0.09 & 10.83 & 1.99 & 99.0 & 9.95 & 96.1 & No \\
%\hline 
 \hline
0.40 & 44 & 2.12 & 11.50  & 6.10 & 1.92 & 12.58  & 0.76 & 35.8 & 6.33 & 55.0&Yes\\
 %\hline
 0.40& 60 &  2.11 & 11.44  & 6.33 & 1.86 &12.74  & 0.89 & 42.2 & 6.73 & 58.8 &Yes\\
 % \hline
 0.40& 76& 2.09 & 11.44 & 6.35 & 1.62  & 12.99 & 1.17& 55.9  & 7.66 & 66.9 & Yes \\
 %\hline 
0.40&92 & 2.08 &11.47  & 6.31 & 1.23 & 13.24 & 1.58  & 75.9 & 8.89 & 77.5 & Yes\\
%\hline
0.40& 100 & 2.04  & 10.82 & 6.64 & 0.55 & 11.87 & 1.91 & 93.6 & 9.78  & 90.3 & No\\
%\hline 
0.40& 108 &  2.03 & 10.64 & 6.60 & 0.27  & 11.23 & 1.99 & 98.0 &  10.01 & 94.0 & No\\
%\hline 
0.40& 116 &  2.03 & 10.51 & 6.59 & 0.10  & 10.95 & 2.01 & 99.0 & 10.07 & 95.8 & No \\
%\hline 
 \hline
0.42 & 44 &2.16 & 11.64  & 5.89 &2.05 &12.58    & 0.47& 21.7 & 5.28 & 45.3&Yes \\
 %\hline
 0.42& 60 &  2.15 & 11.61  & 5.87 & 2.02 & 12.74 & 0.54 & 25.1 &5.55 & 47.8 &Yes\\
  %\hline
 0.42& 76& 2.13 & 11.70 &5.86  & 1.91 & 12.99 & 0.73& 34.2  & 6.30 & 53.8 & Yes\\
 %\hline
0.42&92 & 2.10 & 11.64 &6.04  & 1.41 & 13.48 & 1.49 & 70.9 & 8.55 & 73.4  &Yes \\
%\hline
0.42& 100 & 2.05  & 11.03 &6.37 &0.62 & 12.04 & 1.90 & 92.6   &   9.77& 88.5 & No \\
%\hline 
0.42& 108 &  2.04 & 10.70 & 6.59 & 0.24 & 11.34& 1.99 & 97.5 & 10.00 & 93.4  &No \\
%\hline 
0.42& 116 & 2.04  & 10.56 & 6.58 & 0.11 & 11.00 & 2.01  & 98.5 & 10.07 & 95.3 & No \\
\toprule 
\end{tabular}
\caption{ Some of the neutron stars' main properties.} 
\label{T4}
\end{center}
\end{table}
\end{center}
\end{widetext}

f

We now estimate the mass and size of the quark cores present in the most massive hybrid star of each model
presented. To accomplish that, we follow ref.~\cite{lopesnpa} and solve the TOV equations~\cite{TOV} for the quark EoS from the density corresponding to the central density at the maximum mass hybrid stars and stop at the density corresponding to the critical chemical potential. The percentage amount of quarks present in the most massive hybrid star of each model is presented in Fig.~\ref{Fmq}. As can be seen, as we increase the slope, we also increase the amount of the quark in the most massive hybrid star. For extreme cases ($L = 116$ MeV), the percentual amount of quarks can be as high as $99\%$ of the total mass of the star ($M_{Q(\%)}$), as well 96$\%$ of its radius ($R_{Q(\%)}$), and it is essentially independent of $G_V$. If we restrict ourselves to %only 
models 
%able to fulfill the bonds
that satisfy all the constraints, then we have the amount of quark in both mass and radius around $81\%$, for $L = 92$ MeV. For lower values of $L$, the results strongly depend on $G_V$, and are ultimately related to the critical chemical potential. Nevertheless, we always obtain the total amount of quarks corresponding at least to $20\%$ of the mass of the most massive hybrid star. This fact is in agreement with the discussion presented in ref.~\cite{nature_2020}, about the existence of quarks in the core of massive stars. All the hybrid stars' main results are presented in Tab.~\ref{T4}.

It is worth mentioning a completely different approach, where a
Bayesian analysis of neutron
star properties and some nuclei properties, such as binding energy, charge radii, and neutron skin thickness, were performed, practically ruling out models with very large slopes ~\cite{Mondal:2022cva}. These results are corroborated by our analysis and disfavor extremely large quark cores in hybrid stars. However, we have to keep in mind that the size of the quark core is indeed model-dependent, as claimed above and shown in \cite{Lopes_2022}: while the vector MIT bag model can produce very large cores, the same does not happen if the vector NJL model is used, a feature also obtained in ref.~\cite{Pfaff2022}.

\section{Conclusions}

Motivated by the conflict between the CREX group and the PREX-II results presented in ref.~\cite{Tagami2022}, in this work, we studied how the change in the slope of the symmetry energy at saturation density is correlated to the deconfinement phase transition. For the hadron phase, we used a modified version of QHD including either an $\omega\rho$ interaction to reduce the slope or a $\delta$-meson to increase it. For the quark phase, we used a thermodynamic consistent vector MIT bag model with different quark-meson couplings. 
The main results can be summarized as:

\begin{itemize}
    \item The critical pressure ($p_0$) and the critical chemical potential ($\mu_0$) are strongly dependent of the slope $L$. Lower values of $L$ result in a higher values of $p_0$ and $\mu_0$, while higher values $L$ produce lower values of $p_0$ and $\mu_0$.

    \item Lower values of $G_V$ produces lower values of $p_0$ and $\mu_0$. However, for the higher values of the slope, $L~\geq$  100 MeV, the results are almost independent of $G_V$. Moreover, for $L~\geq$  100 MeV, the critical chemical potential drops below 1050 MeV, which is in disagreement with the discussion pointed out in ref.~\cite{lopesps2}. These results disfavor higher values of the slope, as well as advocate against the scalar-isovector $\delta$ meson.

    \item For low values of $L$, the energy density gap, $\Delta \epsilon$, increases with $G_V$, lying in the range 25 MeV/fm$^3$ $<~\Delta \epsilon~<$ 70 MeV/fm$^3$.  However, as the slope increases, the discontinuity of the energy density quickly grows and becomes almost independent of $G_V$, surpassing 200 MeV/fm$^3$.

    \item A similar behavior can be seen for the latent heat $\Delta Q$. For low values of $L$, it strongly depends on $G_V$. However, as the slope grows, the results are almost independent of $G_V$. $\Delta Q$ presents a local maximum at $L = 100$ MeV, and then quickly drops due to the fast decrease of the critical pressure.

    \item There is also a decrease in the number density on which the phase transition occurs, $n_c$. For $L~\approx~100$ MeV, the number density drops below 1.3 $n_0$. As for $L~\geq~108$ MeV, it drops below the saturation density itself. These results disfavor both, the large values of $L$, as well the use of the scalar-isovector $\delta$ meson.

    \item From the macroscopic point of view, we see that an increase in the slope causes a decrease in the maximum mass. On the other hand, an increase of $G_V$ increases the maximum mass of the hybrid star. Moreover, as the slope $L$ increases, the minimum mass that presents a quark core becomes as low as 0.1 $M_\odot$. For larger values of $L$, both the radius of the canonical star, and the radius related to the PSR J0740+6620 are below the inferior limit constrained by the NICER X-ray telescope~\cite{Riley2021,Riley:2019yda,Miller:2019cac}. This result again disfavors the larger values of $L$ and the use of the $\delta$ meson.

    \item Finally we analyze the total amount of quarks in the most massive hybrid star predicted for each model. For low values of $L$, the results strongly depend on $G_V$. The larger the $G_V$, the lower the total amount of quarks. As the slope increases, the amount of quarks also increases, and the results become almost independent of $G_V$. For extreme values of $L$, the total quark core mass can represent 99\% of the total mass of the hybrid star. However, assuming only parametrizations able to reproduce the constraints, the total mass reaches 81$\%$.

\end{itemize}

\textbf{Acknowledgements:} This work is a part of the project INCT-FNA Proc. No. 464898/2014-5. L.L.L. was partially supported by CNPq Universal Grant No. 409029/2021-1. D.P.M. was partially supported by Conselho Nacional de Desenvolvimento Científico e Tecnológico (CNPq/Brazil) under grant 303490-2021-7.

%%%%%%%%%%%%%%%%%%%%%%%%%
\bibliography{abref}
%%%%%%%%%%%%%%%%%%%%%%%%%%%%%

\end{document}